# Comparative Evaluation of Radiomics and Deep Learning Models for Disease Detection in Chest Radiography

Zhijin He[1,2], Alan B. McMillan[2]

[1]Department of Statistics, [2]Department of Radiology, University of Wisconsin-Madison

## Abstract

The application of artificial intelligence (AI) in medical imaging has revolutionized diagnostic practices, enabling advanced analysis and interpretation of radiological data. This study presents a comprehensive evaluation of radiomics-based and deep learning-based approaches for disease detection in chest radiography, focusing on COVID-19, lung opacity, and viral pneumonia. While deep learning models, particularly convolutional neural networks (CNNs) and vision transformers (ViTs), learn directly from image data, radiomics-based models extract handcrafted features, offering potential advantages in data-limited scenarios. We systematically compared the diagnostic performance of various AI models, including Decision Trees, Gradient Boosting, Random Forests, Support Vector Machines (SVMs), and Multi-Layer Perceptrons (MLPs) for radiomics, against state-of-the-art deep learning models such as InceptionV3, EfficientNetL, and ConvNeXtXLarge. Performance was evaluated across multiple sample sizes. At 24 samples, EfficientNetL achieved an AUC of 0.839, outperforming SVM (AUC = 0.762). At 4000 samples, InceptionV3 achieved the highest AUC of 0.996, compared to 0.885 for Random Forest. A Scheirer–Ray–Hare test confirmed significant main and interaction effects of model type and sample size on all metrics. Post hoc Mann–Whitney U tests with Bonferroni correction further revealed consistent performance advantages for deep learning models across most conditions. These findings provide statistically validated, data-driven recommendations for model selection in diagnostic AI. Deep learning models demonstrated higher performance and better scalability with increasing data availability, while radiomics-based models may remain useful in low-data contexts. This study addresses a critical gap in AI-based diagnostic research by offering practical guidance for deploying AI models across diverse clinical environments.

## Introduction

The application of artificial intelligence (AI) in medical imaging has catalyzed transformative changes in the analysis and interpretation of radiological data, leading to the development of increasingly sophisticated diagnostic tools [1]. Medical images, containing both anatomical and functional details, offer a rich source of information that AI can exploit to enhance diagnostic accuracy, expedite clinical decision-making, and improve patient outcomes [2], [3]. Two principal AI-driven approaches have gained prominence in this field: radiomics-based models and deep learning-based models. Radiomics involves the extraction of quantitative features from images [4], identifying complex patterns that may be beyond human perception, while deep learning-based methods, particularly convolutional neural networks (CNNs) and vision transformers (ViTs), automatically learn hierarchical representations from raw image data without manual feature engineering. Although both methods rely on imaging data, they differ fundamentally in how they approach feature extraction, with each offering unique strengths and limitations in clinical applications.

As the use of AI in diagnostics expands, understanding the comparative performance of radiomics-based and deep learning-based models under various conditions is increasingly critical. Deep learning-based models have demonstrated remarkable scalability and effectiveness when trained on large datasets; however, their performance in settings with limited data remains less certain. In contrast, radiomics-based models, which utilize handcrafted features, may be more resilient to smaller sample sizes and offer potential advantages in data-constrained scenarios [5], [6].

Recent studies have employed AI to enhance the detection and classification of respiratory diseases through radiological diagnostics. Zhang et al. [7] developed a CV19-Net to distinguish COVID-19 pneumonia from other types of pneumonia using chest radiographs. Hu et al. [8] developed radiomics-boosted deep learning models, utilizing a 2D sliding kernel to map radiomic features across chest x-rays, which yielded significantly improved sensitivity and specificity with CNNs like Very Deep Convolutional Networks (VGG) and DenseNet. Kim [9] used Support Vector Machines (SVMs) and Gradient Boosting leveraging radiomic features for more precise COVID-19 and pneumonia classification, with AUC improvements reaching 0.95. Khan et al. [10] proposed a novel deep learning and explainable AI approach. Safari et al. [11] introduced the FuzzyWOA algorithm to optimize the training of deep convolutional neural networks. These studies collectively illustrate the vast potential and current capabilities of AI in enhancing the detection and classification of diseases from radiograph studies, aligning with the goals of our comprehensive evaluation.

However, while prior research has demonstrated the effectiveness of both radiomics-based and deep learning-based approaches independently, there is a lack of systematic, head-to-head comparison of these methods under varying dataset sizes. Specifically, previous studies have not thoroughly investigated how the performance of each approach changes in data-abundant versus data-scarce environments, a crucial consideration for real-world deployment where data availability often varies. This gap necessitates a rigorous evaluation to determine which approach is more suitable under specific clinical constraints. Additionally, existing models often rely on small or imbalanced datasets, which can lead to overfitting and limit generalizability in diverse clinical settings. Few studies have explored how model performance scales across different sample sizes, leaving practitioners with little guidance on method selection under data constraints [12]. The early and accurate detection of these conditions is crucial for timely intervention and improved patient outcomes. Despite its widespread use, manual interpretation of chest radiographs is both labor-intensive and subject to interobserver variability, highlighting the need for reliable, automated diagnostic systems [13], [14]. This retrospective study provides a comprehensive evaluation of dataset size in radiomics-based and deep learning-based approaches in the context of chest radiography for the early detection of lung diseases, including COVID-19, lung opacity, and viral pneumonia. By systematically comparing these approaches across controlled dataset sizes, our work helps bridge this gap and supports more informed decision-making for AI deployment in healthcare. We include a variety of classical machine learning models in our radiomics-based pipeline, such as Decision Tree, which uses a tree-like structure to model decisions and their possible consequences [15]; Random Forest, which builds multiple decision trees and aggregate their predictions to improve generalization [16]; SVM, which finds optimal hyperplanes for classification tasks in high-dimensional space [17]; Gradient Boosting, an ensemble method that sequentially improves weak learners by minimizing errors [18]; and Multi-Layer Perceptron (MLP), which are fully connected feedforward neural networks capable of modeling complex nonlinear relationships [19]. For the deep learning pipeline, we implement several advanced convolutional architectures: ConvNeXtXLarge, a modernized convolutional neural network inspired by the design principles of vision transformers, known for its strong performance on large-scale vision tasks [20]; EfficientNetL, which uses a compound scaling method to balance model depth, width, and resolution for optimized performance and efficiency [21]; and InceptionV3, which uses factorized convolutions and auxiliary classifiers to reduce computational cost and improve gradient flow [22]. These models are trained end-to-end, learning hierarchical features directly from raw pixel data without the need for manual feature engineering. By evaluating their performance across various sample sizes, this study assesses the diagnostic accuracy of each approach. The findings offer valuable insights into the optimal use of AI models for disease detection in chest radiography, particularly in settings with diverse data constraints, and will inform future development of AI-based diagnostic tools for clinical practice.

## Methods

This study followed a multi-stage pipeline involving radiomics feature extraction and data preprocessing, model training, and evaluation. The workflow integrated both radiomics-based and deep learning-based approaches for multi-class classification of chest X-ray images. A visual overview of the methodology is provided in Figure 1.

### Data Source

This study utilized a publicly available and comprehensive dataset of chest X-ray images, sourced from multiple repositories and compiled by a team of international researchers [23]. The dataset contains a total of 21,165 images categorized into four classes: 3,616 images of patients diagnosed with COVID-19, 6,012 images of patients with Lung Opacity, 1,345 images of patients with Viral Pneumonia, and 10,192 images of patients with no disease findings (Normal). The dataset is in the Portable Network Graphics (PNG) format with a resolution of 299×299 pixels and includes a posteroanterior view chest radiograph and its corresponding lung segmentation masks.

The full aggregated dataset was partitioned into training, validation, and testing subsets. A total of 344 samples per class were reserved for the test set using stratified sampling to ensure balanced representation across all four categories. This number was chosen based on the smallest class size (1,345 samples), allowing up to 1,000 samples to remain available for training. The remaining data was used to create training subsets of varying sizes (24, 48, 100, 248, 500, 1,000, 2,000, and 4,000) using stratified sampling to preserve class distribution. A 5-fold cross-validation strategy was employed , with each fold further split into 80% training and 20% validation.

Radiomics Feature Extraction and Preprocessing

A pipeline was employed to prepare medical images and their corresponding lung segmentation masks for radiomics feature extraction, ensuring standardized model inputs. File paths were generated for each subject, systematically organizing image and mask files based on predefined directory structures. Using the SimpleITK [24] library, images and masks were loaded, converted to grayscale if necessary, and rescaled to an intensity range of 0 to 255 using the *RescaleIntensityImageFilter*. Both images and masks were cast to 8-bit unsigned integers to ensure compatibility with subsequent radiomics feature extraction processes. The mask was resampled to match the image size and spatial resolution using nearest-neighbor interpolation, ensuring consistent spatial dimensions and intensity values across inputs.

Radiomic features were extracted using PyRadiomics [25], encompassing both texture-based and intensity-based descriptors. To identify the most informative features for classification, we applied the SelectKBest method from Scikit-learn [26] which ranks features based on the F-values derived from Analysis of Variance (ANOVA). Features such as Gray Level Co-occurrence Matrix (GLCM) Cluster Shade and Cluster Tendency (which quantify texture heterogeneity), along with First-order Skewness and Interquartile Range (which describe intensity distribution), were among the highest-ranked based on their ANOVA F-scores. The data was normalized, and reference standard annotations (0 for normal, 1 for COVID, 2 for viral pneumonia, and 3 for lung opacity) were assigned to each class.

Deep Learning-Based Model Data Preprocessing

For the image-based deep learning models, the data was preprocessed to match the input format required by the models. Images were resized to 256x256 pixels, and pixel values were normalized to the [0, 1] range. Data augmentation was performed using TensorFlow [27]'s *ImageDataGenerator*, including rescaling, shear transformations, zoom operations, and horizontal flipping to simulate variability in chest X-ray images. This preprocessing ensured that model inputs were consistent with the preprocessed image format and enhanced the model's generalization capabilities. The images were organized into directory structures for efficient loading via the *flow_from_directory* method.

Radiomics-Based Model Training

For radiomics-based classification, both traditional machine learning algorithms and a deep learning model were implemented to evaluate performance on handcrafted feature sets. Traditional classifiers included SVM, Decision Trees, Gradient Boosting, and Random Forests, all developed using Scikit-Learn. Input features were standardized using the StandardScaler, and dimensionality reduction was performed using the *SelectKBest* method with ANOVA F-value (*f_classif*) as the scoring function. For each model, subsets of the top *k* features (ranging from 1 to the total number of extracted features) were evaluated to identify the configuration that yielded the highest performance. The SVM classifier (SVC) was configured with probability=True and used the default RBF kernel (*C=1.0, gamma='scale'*). Decision Tree, Gradient Boosting, and Random Forest classifiers were used with default hyperparameters. In addition to these traditional models, a deep learning-based MLP was implemented using Keras/TensorFlow to assess the suitability of neural architectures for radiomics data. The MLP architecture consisted of a 64-unit dense layer, followed by a dropout layer (dropout rate = 0.2), a 32-unit dense layer, and a final softmax output layer for multi-class classification. The MLP was trained for 100 epochs, consistent with common deep learning training practices. Randomized operations such as data splitting, undersampling, feature selection, and model initialization were controlled using a fixed random seed (*random_state=42*). For each model, the configuration yielding the maximum F1 score across all k-feature evaluations was retained for comparative analysis.

Deep Learning-Based Model Training

Image-based deep learning models, including ConvNeXtXLarge, EfficientNetL, and InceptionV3, were employed. These models utilized architectures pre-trained on ImageNet, with their base weights fine-tuned end-to-end for the specific classification task. TensorFlow/Keras [28] was used to build sequential models, with a consistent classification head added to each base network. This included a global average pooling layer, a 256-unit dense layer with ReLU activation, batch normalization, a 0.5 dropout layer to prevent overfitting, and a final softmax output layer to produce class probabilities for multi-class classification. All deep learning-based models were trained using the Adam optimizer with a learning rate of 0.0001 and a batch size of 16. Training was conducted for 100 epochs. The loss function used was categorical cross-entropy. All random operations, including image sampling, fold generation, and weight initialization, were controlled using *random_state=42*.

Model Outputs and Evaluation

The outputs of both the radiomics-based and image-based models were designed to predict the likelihood of each class (normal, COVID-19, viral pneumonia, and lung opacity), in line with the clinical requirement for multi-class classification of chest X-ray images. Model performance was evaluated using metrics such as F1 score, AUC score, accuracy, sensitivity, and specificity, which were averaged across folds and runs to provide a comprehensive assessment of model performance.

The performance of various models across different sample sizes was rigorously analyzed to identify statistically significant differences in these five key metrics. A non-parametric Scheirer-Ray-Hare (SRH) test was employed to provide a statistical understanding of the effect of model selection and sample size [29]. This test, fundamentally an ANOVA performed on ranks, was implemented by converting the original scores for each metric into their respective ranks before applying a standard two-way ANOVA. Furthermore, pairwise Mann-Whitney U tests were conducted to determine specific group differences for sample size and model by pooling across model and sample size, respectively. To mitigate the risk of Type I errors arising from multiple comparisons, p-values from these pairwise tests were adjusted using the Bonferroni correction.

Radiomics-based model training was conducted using Python 3.10.12 with the following libraries and versions: imbalanced-learn 0.13.0, scikit-learn 1.5.2, pandas 2.2.2, NumPy 2.0.2, and TensorFlow 2.17.0. Deep learning-based model training was performed using Python 3.9.23 with the following libraries and versions: scikit-learn 1.6.1, NumPy 2.0.2, SciPy 1.13.1, TensorFlow 2.17.0, joblib 1.4.2, and threadpoolctl 3.5.0. Statistical analyses were conducted using statsmodels 0.14.5 and pingouin 0.5.5. Figures and tables were produced using pandas 2.2.2, Matplotlib 3.8.4, and Seaborn 0.13.2.

## Results

We evaluated the performance of traditional radiomics-based models and deep learning architectures across five classification metrics, F1 score, AUC, accuracy, sensitivity, and specificity, at varying training sample sizes ranging from 24 to 4000. As shown in Table 1, model performance improved consistently with increasing sample size, especially for deep learning-based models. InceptionV3 achieved the highest AUC and accuracy scores, with a peak AUC of 0.996 and accuracy of 0.960 at 4000 samples. EfficientNetL led in small to medium sample sizes, while ConvNeXtXLarge exhibited strong improvements with more data, eventually approaching the performance of the other top-tier deep learning-based models. Among the traditional models, Random Forest and SVM performed relatively well at smaller sample sizes but showed limited improvement beyond 1000 samples. Standard deviations, also reported in Table 1, were higher at smaller sample sizes, particularly for traditional machine learning models like Decision Tree and Gradient Boosting, indicating high variability. These deviations decreased substantially as training size increased, especially for deep learning models. For example, EfficientNetL's standard deviation in F1 score dropped from ±0.042 at 24 samples to ±0.006 at 4000 samples, highlighting the stabilizing effect of larger datasets on complex architectures.

To statistically validate these trends, we performed a non-parametric SRH test, the results of which are presented in Table 2. Here, $p_{srh}$ represents p-values from the SRH test, the H statistic is the chi-square value calculated from ranked data that quantifies how far the average ranks differ among the levels of each factor (and their interaction). The SRH test revealed significant main and interaction effects of model type and sample size across all five metrics. For instance, AUC was strongly influenced by model ($H = 913.47$, $p_{srh} \ll 1\times 10^{-5}$) and sample size ($H = 652.79$, $p_{srh} \ll 1\times 10^{-5}$), with a statistically significant interaction ($H = 9.61$, $p_{srh} \ll 1\times 10^{-5}$), indicating that improvements due to data size were model-dependent. Similar patterns were observed for F1 score ($H = 268.92$, $p_{srh} \ll 1\times 10^{-5}$ for model; $H = 459.72$, $p_{srh} \ll 1\times 10^{-5}$ for sample size), accuracy, specificity, and sensitivity, all of which demonstrated significant main and interaction effects.

We conducted pairwise Mann–Whitney U tests with Bonferroni correction to provide a detailed landscape of performance differences across various models and sample sizes, with models pooled when evaluating sample size effects and sample sizes pooled for model comparisons. As shown in Table 3, in the pairwise evaluations of model performance, substantial and highly statistically significant differences were observed for numerous pairings across all assessed metrics including F1-score, AUC, accuracy, specificity, and sensitivity. For example, the comparison between ConvNeXtXLarge and Decision Tree consistently revealed extremely low Bonferroni corrected p-values across all metrics (e.g., F1 score: $6.73 \times 10^{-15}$ , AUC: $1.31 \times 10^{-33}$ , accuracy: $3.77 \times 10^{-17}$ ), indicating that these models perform differently. Similarly, ConvNeXtXLarge also demonstrated highly significant differences when compared against Random Forest, with p-values of $5.89 \times 10^{-11}$ for F1 score, $1.56 \times 10^{-17}$ for AUC, and $3.31 \times 10^{-13}$ for accuracy. In contrast, some model pairings exhibited no statistically significant performance difference after Bonferroni correction, such as comparing ConvNeXtXLarge against both EfficientNetL and InceptionV3, suggesting these models perform equivalently under the tested conditions. For the analysis of varying sample sizes, where models were pooled, there was a pronounced and consistent impact of sample size on all performance metrics. As the difference in sample size between compared groups increased, the p-values for all metrics decreased dramatically, signifying increasingly significant performance variations. For instance, the comparison of 24 samples versus 2000 samples yielded exceptionally low p-values across the board, such as $5.48 \times 10^{-21}$ for F1 score, $3.21 \times 10^{-18}$ for AUC, and $3.49\times10^{-19}$ for accuracy. Even moderate increases in sample size demonstrated significant impacts; for example, comparing 24 samples against 100 samples resulted in p-values of $3.01\times10^{-13}$ for F1 score, $1.33 \times 10^{-5}$ for AUC, and $7.08 \times 10^{-10}$ for accuracy.

The effect of training sample size was visualized in Figure 2, which illustrated upward trends in all metrics as data volume increased. While models like SVM and Random Forest improved steadily (e.g., Random Forest AUC increased from 0.741 at 24 samples to 0.886 at 2000), their final performance at high sample sizes remained significantly below that of deep learning-based models. In contrast, InceptionV3's AUC rose from 0.844 at 24 samples to 0.996 at 4000, mirroring gains across all other metrics. EfficientNetL and ConvNeXtXLarge followed similar trajectories, confirming their scalability and diagnostic accuracy with larger datasets.

**Discussion**

This study evaluated the performance of radiomics-based and deep learning-based models for classification using datasets of varying sample sizes. The models tested included SVM, Random Forest, Gradient Boosting, Decision Tree, MLP, InceptionV3, EfficientNetL, and ConvNeXtXLarge. Model performance was assessed using F1 score, AUC score, accuracy, sensitivity, and specificity. Larger sample sizes generally led to improved model performance, with deep learning-based models benefiting more. At the largest sample size of 4000, InceptionV3 achieved an AUC score of 0.996 and accuracy of 0.960, outperforming other models. EfficientNetL followed closely with an AUC score of 0.994, showing similar improvements with increasing data. These results indicate that deep learning-based models generalize better with more data, reducing performance variability.

Among traditional models, Random Forest and SVM showed relatively strong performance, especially at larger sample sizes. By 4000 samples, both reached AUC scores close to 0.88, suggesting that traditional models remain viable when sufficient data is available. MLP demonstrated the most notable improvement among traditional models, with performance gains especially visible in F1 score and AUC at higher sample sizes. In contrast, Decision Tree and Gradient Boosting consistently underperformed, particularly in smaller datasets, where their performance and stability were lowest. These results suggest that simpler models may be less effective in complex or noisy classification tasks, especially when data is limited.

Smaller sample sizes led to greater variability in model outputs across the board. Standard deviations were highest at 24 and 48 samples, reflecting unstable learning behavior. Even under these conditions, deep learning-based models like InceptionV3 and EfficientNetL still outperformed traditional models, although with elevated variance. As sample size increased, the performance gap widened, and deep learning-based models not only improved more rapidly but also demonstrated greater consistency, as evidenced by decreasing standard deviations in Table 1. On the other hand, Random Forest and SVM showed more stable but modest gains, making them reliable choices when training data is constrained.

To validate these observations, we utilized the non-parametric SRH test, which confirmed that both model architecture and training data volume had statistically significant main effects, along with significant interaction effects across all performance metrics ($p_{srh} \ll 1\times 10^{-5}$; Table 2). This non-parametric approach was well-suited for our data, as it does not assume normality or homogeneity of variances. The corrected pairwise Mann-Whitney U tests (Table 3) provide insight into the comparative performance of models and the influence of sample size, thereby reinforcing the overall findings from the SRH test. For example, there were consistent performance differences between the ConvNeXtXLarge and Decision Tree models, and statistically equivalent performance between models such as InceptionV3 and EfficientNetL. Furthermore, the impact of increasing sample size across all performance metrics underscores the direct relationship between the amount of data and model performance. This confirms that larger datasets are crucial not only for training better performing models but also for revealing subtle yet significant performance variations that might be obscured in smaller cohorts, thereby informing future data collection strategies and model development priorities.

Model selection, therefore, should be guided by data availability and computational constraints. When large datasets are accessible, deep learning-based models such as InceptionV3, EfficientNetL, and ConvNeXtXLarge provide the best performance, with high AUC and accuracy metrics, and reduced variance at scale. However, their complexity and lack of interpretability may pose challenges in settings where transparency or computational efficiency is required. Traditional models like Random Forest and SVM offer more interpretable alternatives with respectable performance, especially at moderate data sizes. MLP may be a particularly promising choice among traditional models when working with datasets in the 1000–4000 sample range. Conversely, Decision Tree and Gradient Boosting were comparatively unreliable, especially at low sample sizes, and may not be suitable for real-world diagnostic applications where data scarcity is common.

In comparison with prior work, our models outperformed several published benchmarks, despite differences in datasets and sample sizes as shown in Table 4. For example, EfficientNetL achieved an AUC of 0.972 with 500 samples and 0.949 with 248 samples, outperforming Zhang et al.'s CV19-Net (AUC = 0.94) [7], which was evaluated on a different dataset. At 4000 samples, InceptionV3 reached an AUC of 0.996, exceeding Hu et al.'s VGG-19 (AUC = 0.987 on 812 images) [8] and Khan et al.'s hybrid model combining EfficientNet, VGG16, and ELM (AUC = 0.991 on 34,579 samples) [10]. Note that Khan et al. utilized the same dataset used herein, used herein, alongside additional sources. Our deep learning-based models also demonstrated more effective scalability compared to Paul et al.'s ensemble of autoencoders (AUC = 0.79 on 25 samples) [30], and delivered competitive performance relative to Chaddad et al.'s radiomics approach (AUC = 0.9945 on 5254 samples) [31], though dataset differences may limit direct comparison.

This study has some limitations. First, the models were trained and tested on a single dataset, limiting their generalizability to external or out-of-distribution data. Although our statistical results were robust, they may still be influenced by dataset-specific biases such as imaging protocols or patient demographics. Validation across independent datasets from different clinical settings is needed to confirm the broader applicability of our models. Additionally, variation in image acquisition and quality may impact model performance, especially in multi-center applications. Another limitation is the relatively simple radiomics-based feature extraction used in our traditional models. Future work could explore more advanced radiomics pipelines or hybrid frameworks that integrate radiomic and image-based features, particularly for cases where both structured and unstructured data are available.

**Conclusion**

This study highlights the strong performance of deep learning-based models, particularly InceptionV3 and EfficientNetL, in classifying medical images. These models performed best with larger datasets, demonstrating their ability to generalize well and handle complex tasks. For settings with large datasets, deep learning-based models are the ideal choice due to their superior performance. However, for healthcare providers with smaller datasets, traditional models like SVM and Random Forest remain reliable and efficient alternatives, offering stable performance even with limited data. The use of both parametric and non-parametric statistical analyses confirmed that model selection and dataset size interact significantly, reinforcing the validity of the observed performance differences. The results emphasize the importance of tailoring model selection to the available data and resources in different healthcare environments, ensuring that each institution can select the most effective model for their specific needs.

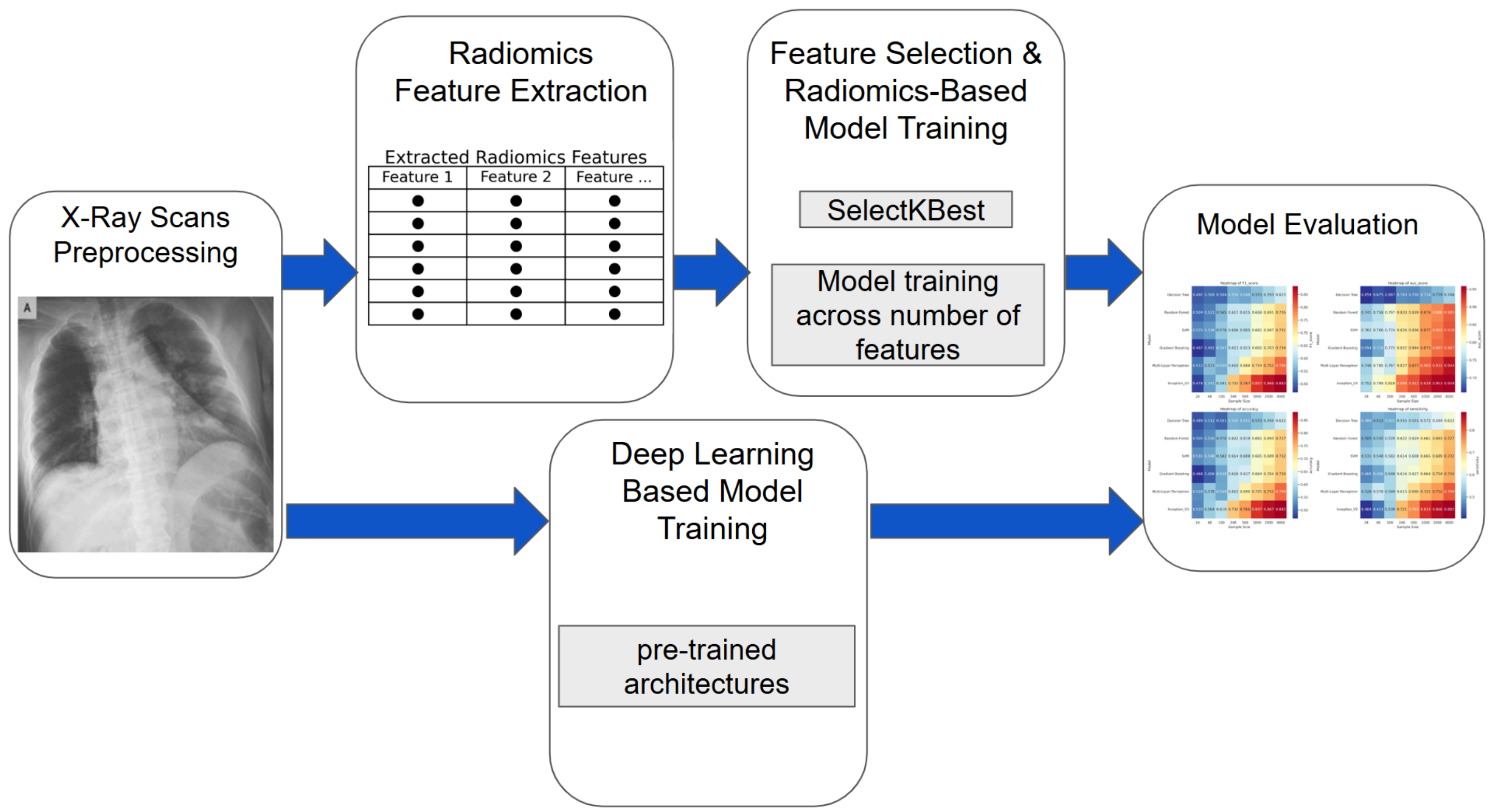

**Figure 1.** Overview of workflow for this study. The workflow integrates radiomics and deep learning for X-ray analysis. It starts with X-ray preprocessing, followed by radiomics feature extraction. Feature selection (SelectKBest) and model training are performed, while deep learning models are trained separately. Finally, both approaches undergo model evaluation using performance metrics.

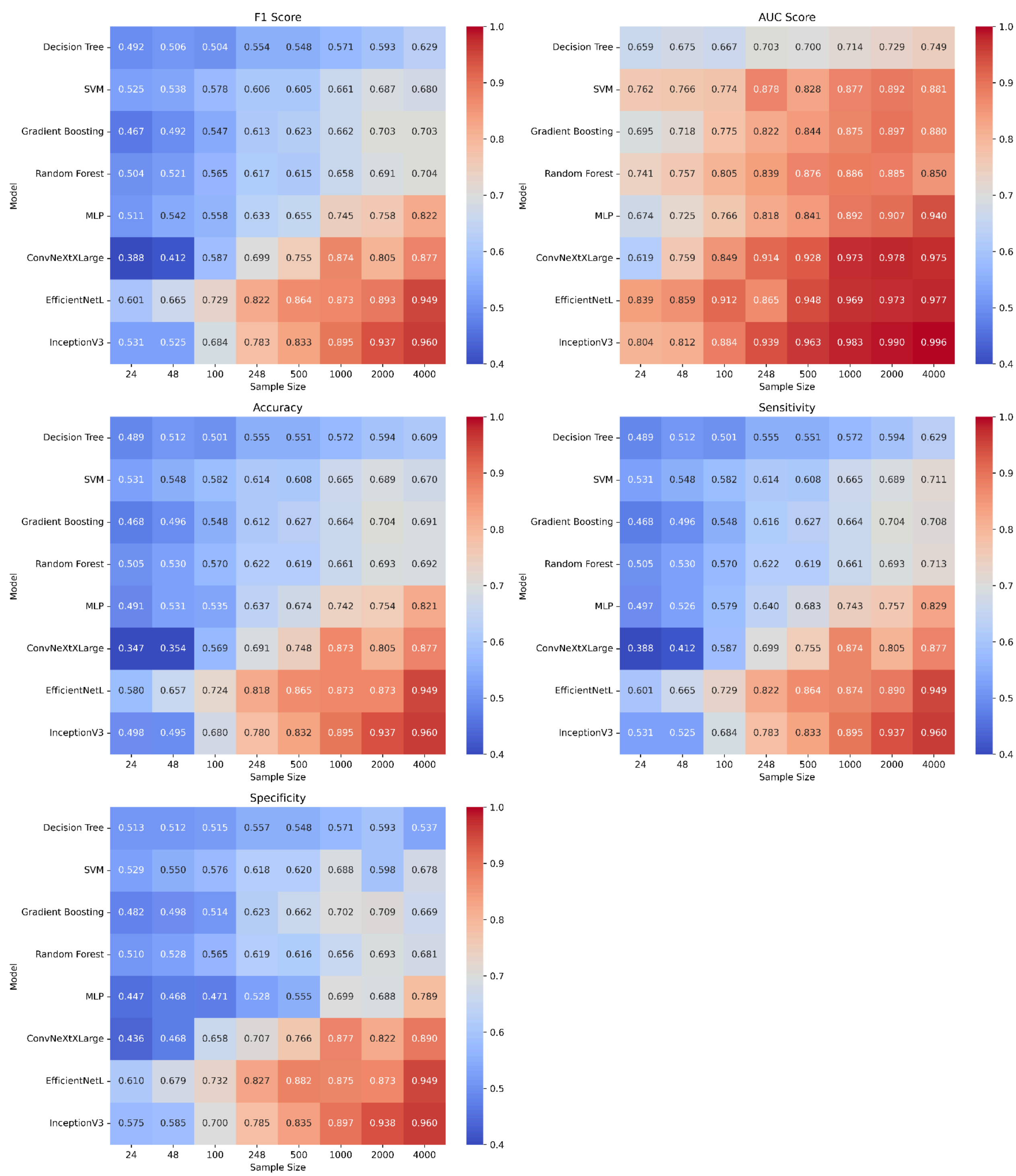

**Figure 2.** Heatmaps of model performance across key metrics and a range of sample sizes. The heatmaps display F1 Score, AUC Score, Accuracy, Sensitivity, and Specificity of the models (Decision Tree, SVM, Gradient Boosting, Random Forest, MLP, ConvNeXtXLarge, EfficientNetL, and InceptionV3) across sample sizes from 24 to 4000. EfficientNetL shows strong performance at smaller sample sizes (24, 48, 100, 248, 500), while InceptionV3 consistently performs well at larger sample sizes (1000, 2000, 4000), especially in accuracy, AUC, and sensitivity.

| Model | Sample Size | F1 Score | AUC Score | Accuracy | Sensitivity | Specificity |
|---|---|---|---|---|---|---|
| Decision Tree | 24 | 0.492 ± 0.022 | 0.659 ± 0.017 | 0.489 ± 0.025 | 0.489 ± 0.025 | 0.513 ± 0.024 |
| Random Forest | 24 | 0.504 ± 0.030 | 0.741 ± 0.028 | 0.505 ± 0.027 | 0.505 ± 0.027 | 0.510 ± 0.030 |
| Gradient Boosting | 24 | 0.467 ± 0.028 | 0.695 ± 0.025 | 0.468 ± 0.024 | 0.468 ± 0.024 | 0.482 ± 0.028 |
| SVM | 24 | 0.525 ± 0.027 | 0.762 ± 0.025 | 0.531 ± 0.026 | 0.531 ± 0.026 | 0.529 ± 0.027 |
| MLP | 24 | 0.511 ± 0.041 | 0.674 ± 0.074 | 0.491 ± 0.044 | 0.497 ± 0.048 | 0.447 ± 0.044 |
| InceptionV3 | 24 | 0.531 ± 0.052 | 0.804 ± 0.014 | 0.498 ± 0.059 | 0.531 ± 0.052 | 0.575 ± 0.040 |
| **EfficientNetL** | **24** | **0.601 ± 0.042** | **0.839 ± 0.013** | **0.580 ± 0.047** | **0.601 ± 0.042** | **0.610 ± 0.044** |
| ConvNextXLarge | 24 | 0.388 ± 0.062 | 0.691 ± 0.036 | 0.347 ± 0.087 | 0.388 ± 0.062 | 0.436 ± 0.045 |
| Decision Tree | 48 | 0.506 ± 0.021 | 0.675 ± 0.014 | 0.512 ± 0.022 | 0.512 ± 0.022 | 0.512 ± 0.022 |
| Random Forest | 48 | 0.521 ± 0.029 | 0.758 ± 0.025 | 0.530 ± 0.026 | 0.530 ± 0.026 | 0.528 ± 0.029 |
| Gradient Boosting | 48 | 0.492 ± 0.029 | 0.718 ± 0.022 | 0.496 ± 0.022 | 0.496 ± 0.022 | 0.498 ± 0.029 |
| SVM | 48 | 0.538 ± 0.026 | 0.766 ± 0.022 | 0.548 ± 0.025 | 0.548 ± 0.025 | 0.550 ± 0.026 |
| MLP | 48 | 0.542 ± 0.035 | 0.725 ± 0.037 | 0.531 ± 0.042 | 0.526 ± 0.016 | 0.468 ± 0.030 |
| InceptionV3 | 48 | 0.525 ± 0.034 | 0.812 ± 0.027 | 0.495 ± 0.031 | 0.525 ± 0.034 | 0.585 ± 0.045 |
| **EfficientNetL** | **48** | **0.665 ± 0.016** | **0.859 ± 0.009** | **0.657 ± 0.017** | **0.665 ± 0.016** | **0.659 ± 0.018** |
| ConvNexXtXLarge | 48 | 0.412 ± 0.104 | 0.759 ± 0.013 | 0.354 ± 0.158 | 0.412 ± 0.104 | 0.468 ± 0.204 |
| Decision Tree | 100 | 0.504 ± 0.023 | 0.667 ± 0.015 | 0.501 ± 0.021 | 0.501 ± 0.021 | 0.515 ± 0.022 |
| Random Forest | 100 | 0.565 ± 0.027 | 0.797 ± 0.022 | 0.570 ± 0.024 | 0.570 ± 0.024 | 0.565 ± 0.027 |
| Gradient Boosting | 100 | 0.547 ± 0.027 | 0.775 ± 0.020 | 0.548 ± 0.020 | 0.548 ± 0.020 | 0.551 ± 0.027 |
| SVM | 100 | 0.578 ± 0.024 | 0.774 ± 0.020 | 0.582 ± 0.023 | 0.582 ± 0.023 | 0.576 ± 0.024 |
| MLP | 100 | 0.558 ± 0.020 | 0.766 ± 0.007 | 0.535 ± 0.015 | 0.579 ± 0.019 | 0.471 ± 0.007 |
| InceptionV3 | 100 | 0.684 ± 0.023 | 0.889 ± 0.004 | 0.680 ± 0.026 | 0.684 ± 0.023 | 0.700 ± 0.019 |
| **EfficientNetL** | **100** | **0.729 ± 0.040** | **0.902 ± 0.023** | **0.724 ± 0.042** | **0.729 ± 0.040** | **0.732 ± 0.042** |
| ConvNeXtXLarge | 100 | 0.587 ± 0.043 | 0.849 ± 0.026 | 0.569 ± 0.053 | 0.587 ± 0.043 | 0.658 ± 0.033 |
| Decision Tree | 248 | 0.554 ± 0.019 | 0.703 ± 0.016 | 0.555 ± 0.019 | 0.555 ± 0.019 | 0.557 ± 0.020 |
| Random Forest | 248 | 0.617 ± 0.025 | 0.833 ± 0.020 | 0.622 ± 0.021 | 0.622 ± 0.021 | 0.619 ± 0.025 |
| Gradient Boosting | 248 | 0.613 ± 0.024 | 0.822 ± 0.018 | 0.616 ± 0.019 | 0.616 ± 0.019 | 0.614 ± 0.024 |
| SVM | 248 | 0.606 ± 0.022 | 0.878 ± 0.018 | 0.614 ± 0.021 | 0.614 ± 0.021 | 0.618 ± 0.022 |
| MLP | 248 | 0.633 ± 0.019 | 0.818 ± 0.004 | 0.613 ± 0.022 | 0.640 ± 0.019 | 0.528 ± 0.013 |
| InceptionV3 | 248 | 0.783 ± 0.030 | 0.939 ± 0.012 | 0.780 ± 0.030 | 0.783 ± 0.030 | 0.785 ± 0.033 |
| **EfficientNetL** | **248** | **0.822 ± 0.032** | **0.948 ± 0.012** | **0.818 ± 0.034** | **0.822 ± 0.032** | **0.827 ± 0.029** |
| ConvNextXLarge | 248 | 0.699 ± 0.035 | 0.900 ± 0.013 | 0.691 ± 0.043 | 0.699 ± 0.035 | 0.707 ± 0.016 |
| Decision Tree | 500 | 0.548 ± 0.022 | 0.700 ± 0.015 | 0.551 ± 0.020 | 0.551 ± 0.020 | 0.548 ± 0.021 |
| Random Forest | 500 | 0.615 ± 0.024 | 0.839 ± 0.019 | 0.619 ± 0.022 | 0.619 ± 0.022 | 0.616 ± 0.024 |
| Gradient Boosting | 500 | 0.623 ± 0.021 | 0.844 ± 0.019 | 0.627 ± 0.021 | 0.627 ± 0.021 | 0.623 ± 0.021 |
| SVM | 500 | 0.605 ± 0.023 | 0.828 ± 0.019 | 0.608 ± 0.022 | 0.608 ± 0.022 | 0.612 ± 0.023 |
| MLP | 500 | 0.655 ± 0.020 | 0.841 ± 0.096 | 0.637 ± 0.214 | 0.683 ± 0.013 | 0.555 ± 0.016 |
| InceptionV3 | 500 | 0.833 ± 0.016 | 0.963 ± 0.005 | 0.832 ± 0.016 | 0.833 ± 0.016 | 0.835 ± 0.015 |
| **EfficientNetL** | **500** | **0.864 ± 0.015** | **0.969 ± 0.002** | **0.865 ± 0.015** | **0.864 ± 0.015** | **0.872 ± 0.012** |
| ConvNeXtXLarge | 500 | 0.755 ± 0.028 | 0.928 ± 0.014 | 0.748 ± 0.030 | 0.755 ± 0.028 | 0.766 ± 0.021 |
| Decision Tree | 1000 | 0.571 ± 0.021 | 0.714 ± 0.013 | 0.572 ± 0.019 | 0.661 ± 0.019 | 0.571 ± 0.020 |
| Random Forest | 1000 | 0.658 ± 0.021 | 0.876 ± 0.018 | 0.661 ± 0.019 | 0.664 ± 0.020 | 0.658 ± 0.021 |
| Gradient Boosting | 1000 | 0.662 ± 0.023 | 0.875 ± 0.017 | 0.664 ± 0.020 | 0.665 ± 0.019 | 0.662 ± 0.023 |
| SVM | 1000 | 0.661 ± 0.020 | 0.877 ± 0.018 | 0.665 ± 0.019 | 0.743 ± 0.008 | 0.670 ± 0.020 |
| MLP | 1000 | 0.745 ± 0.005 | 0.892 ± 0.006 | 0.742 ± 0.005 | 0.895 ± 0.025 | 0.699 ± 0.008 |
| **InceptionV3** | **1000** | **0.895 ± 0.025** | **0.983 ± 0.005** | **0.895 ± 0.025** | **0.874 ± 0.038** | **0.897 ± 0.024** |
| EfficientNetL | 1000 | 0.874 ± 0.038 | 0.973 ± 0.008 | 0.873 ± 0.039 | 0.874 ± 0.042 | 0.878 ± 0.033 |
| ConvNexXtXLarge | 1000 | 0.874 ± 0.042 | 0.973 ± 0.009 | 0.873 ± 0.043 | 0.594 ± 0.019 | 0.877 ± 0.036 |
| Decision Tree | 2000 | 0.593 ± 0.019 | 0.729 ± 0.013 | 0.594 ± 0.019 | 0.693 ± 0.018 | 0.593 ± 0.019 |
| Random Forest | 2000 | 0.691 ± 0.019 | 0.886 ± 0.017 | 0.693 ± 0.018 | 0.704 ± 0.019 | 0.691 ± 0.019 |
| Gradient Boosting | 2000 | 0.703 ± 0.020 | 0.897 ± 0.015 | 0.704 ± 0.019 | 0.689 ± 0.018 | 0.702 ± 0.020 |
| SVM | 2000 | 0.687 ± 0.019 | 0.892 ± 0.017 | 0.689 ± 0.018 | 0.757 ± 0.004 | 0.688 ± 0.019 |
| MLP | 2000 | 0.758 ± 0.003 | 0.907 ± 0.006 | 0.754 ± 0.003 | 0.937 ± 0.006 | 0.688 ± 0.008 |
| **InceptionV3** | **2000** | **0.937 ± 0.006** | **0.990 ± 0.003** | **0.937 ± 0.006** | **0.880 ± 0.075** | **0.938 ± 0.006** |
| EfficientNetL | 2000 | 0.880 ± 0.075 | 0.977 ± 0.019 | 0.873 ± 0.085 | 0.805 ± 0.019 | 0.893 ± 0.058 |
| ConvNextXLarge | 2000 | 0.805 ± 0.019 | 0.957 ± 0.009 | 0.805 ± 0.019 | 0.629 ± 0.026 | 0.822 ± 0.009 |
| Decision Tree | 4000 | 0.629 ± 0.018 | 0.749 ± 0.016 | 0.609 ± 0.021 | 0.713 ± 0.012 | 0.537 ± 0.020 |
| Random Forest | 4000 | 0.704 ± 0.009 | 0.885 ± 0.003 | 0.692 ± 0.008 | 0.708 ± 0.020 | 0.609 ± 0.008 |
| Gradient Boosting | 4000 | 0.703 ± 0.006 | 0.880 ± 0.003 | 0.691 ± 0.007 | 0.829 ± 0.008 | 0.609 ± 0.010 |
| SVM | 4000 | 0.680 ± 0.003 | 0.881 ± 0.002 | 0.670 ± 0.003 | 0.960 ± 0.008 | 0.598 ± 0.004 |
| MLP | 4000 | 0.822 ± 0.002 | 0.940 ± 0.005 | 0.821 ± 0.001 | 0.949 ± 0.009 | 0.789 ± 0.013 |
| **InceptionV3** | **4000** | **0.960 ± 0.008** | **0.996 ± 0.001** | **0.960 ± 0.008** | **0.877 ± 0.033** | **0.960 ± 0.008** |
| EfficientNetL | 4000 | 0.949 ± 0.009 | 0.994 ± 0.002 | 0.949 ± 0.009 | 0.949 ± 0.009 | 0.949 ± 0.009 |
| ConvNextXLarge | 4000 | 0.875 ± 0.035 | 0.978 ± 0.006 | 0.877 ± 0.033 | 0.877 ± 0.033 | 0.890 ± 0.013 |

**Table 1** Quantitative comparison of performance metrics for the machine learning and deep learning models evaluated across various sample sizes ranging from 24 to 4000. The models include Decision Tree, Random Forest, Gradient Boosting, Support Vector Machine (SVM), Multi-Layer Perceptron (MLP), InceptionV3, EfficientNetL, and ConvNeXtXLarge. Performance is measured using key evaluation metrics: F1 Score, Area Under the Curve (AUC) Score, Accuracy, Sensitivity, and Specificity. All metric values are rounded to three decimal places. EfficientNetL performs best at small and medium sample sizes (24, 48, 100, 248, and 500) while InceptionV3 achieves the highest performance at larger sample sizes (1000, 2000, and 4000), particularly in terms of accuracy, AUC, and sensitivity

| Metric | Factor | H | $p_{srh}$ |
|---|---|---|---|
| F1 | Model | 268.92 | $p \ll 10^{-5}$ |
| F1 | Sample Size | 459.72 | $p \ll 10^{-5}$ |
| F1 | Model:Sample Size | 7.21 | $p \ll 10^{-5}$ |
| AUC | Model | 913.47 | $p \ll 10^{-5}$ |
| AUC | Sample Size | 652.79 | $p \ll 10^{-5}$ |
| AUC | Model:Sample Size | 9.61 | $p \ll 10^{-5}$ |
| Accuracy | Model | 310.43 | $p \ll 10^{-5}$ |
| Accuracy | Sample Size | 399.33 | $p \ll 10^{-5}$ |
| Accuracy | Model:Sample Size | 5.92 | $p \ll 10^{-5}$ |
| Specificity | Model | 525.70 | $p \ll 10^{-5}$ |
| Specificity | Sample Size | 344.69 | $p \ll 10^{-5}$ |
| Specificity | Model:Sample Size | 6.27 | $p \ll 10^{-5}$ |
| Sensitivity | Model | 295.25 | $p \ll 10^{-5}$ |
| Sensitivity | Sample Size | 422.23 | $p \ll 10^{-5}$ |
| Sensitivity | Model:Sample Size | 5.16 | $p \ll 10^{-5}$ |

**Table 2.** Results of the Schierer-Ray-Hare analysis assessing the influence of model type, training sample size, and their interaction on classification performance across five key evaluation metrics: F1 Score, AUC, Accuracy, Sensitivity, and Specificity. For each metric, the Sum of Squares, degree of freedom, F-statistics and corresponding p-values are reported for the main effects (Model and Sample Size) and their interaction (Model × Sample Size). All reported p-values are statistically significant ($p_{srh} \ll 1\times 10^{-5}$), indicating that both model architecture and sample size independently and interactively impact classification performance.

| Contrast | A | B | p-value (F1) | p-value (AUC) | p-value (Accuracy) | p-value (Specificity) | p-value (Sensitivity) |
|---|---|---|---|---|---|---|---|
| Model | ConvNeXtXLarge | Decision Tree | **$6.73\times10^{-15}$** | **$1.31\times10^{-33}$** | **$3.77\times10^{-17}$** | **$5.70\times10^{-20}$** | **$2.78\times10^{-17}$** |
| Model | ConvNeXtXLarge | EfficientNetL | 1.00 | 1.00 | 1.00 | 1.00 | 1.00 |
| Model | ConvNeXtXLarge | Gradient Boosting | **$6.93\times10^{-12}$** | **$5.04\times10^{-17}$** | **$4.92\times10^{-14}$** | **$1.06\times10^{-18}$** | **$2.67\times10^{-13}$** |
| Model | ConvNeXtXLarge | InceptionV3 | 1.00 | 1.00 | 1.00 | 1.00 | 1.00 |
| Model | ConvNeXtXLarge | MLP | **$7.33\times10^{-5}$** | **$7.53\times10^{-9}$** | **$3.26\times10^{-6}$** | **$7.00\times10^{-11}$** | **$2.20\times10^{-5}$** |
| Model | ConvNeXtXLarge | Random Forest | **$5.89\times10^{-11}$** | **$1.56\times10^{-17}$** | **$3.31\times10^{-13}$** | **$1.71\times10^{-17}$** | **$5.66\times10^{-13}$** |
| Model | ConvNeXtXLarge | SVM | **$4.48\times10^{-11}$** | **$4.19\times10^{-13}$** | **$1.72\times10^{-13}$** | **$1.59\times10^{-17}$** | **$8.91\times10^{-13}$** |
| Model | Decision Tree | EfficientNetL | **$6.73\times10^{-15}$** | **$1.31\times10^{-33}$** | **$3.77\times10^{-17}$** | **$5.70\times10^{-20}$** | **$2.78\times10^{-17}$** |
| Model | Decision Tree | Gradient Boosting | $2.26\times10^{-1}$ | **$7.59\times10^{-9}$** | $1.30\times10^{-1}$ | $1.47\times10^{-1}$ | $2.24\times10^{-1}$ |
| Model | Decision Tree | InceptionV3 | **$4.43\times10^{-8}$** | **$1.04\times10^{-24}$** | **$2.57\times10^{-10}$** | **$9.51\times10^{-16}$** | **$2.94\times10^{-10}$** |
| Model | Decision Tree | MLP | **$8.66\times10^{-11}$** | **$1.97\times10^{-18}$** | **$1.77\times10^{-10}$** | **$2.13\times10^{-8}$** | **$2.33\times10^{-11}$** |
| Model | Decision Tree | Random Forest | **$1.90\times10^{-3}$** | **$6.63\times10^{-17}$** | **$2.14\times10^{-3}$** | **$4.07\times10^{-4}$** | **$8.32\times10^{-3}$** |
| Model | Decision Tree | SVM | **$2.44\times10^{-2}$** | **$1.05\times10^{-3}$** | **$2.24\times10^{-2}$** | **$7.33\times10^{-4}$** | **$3.83\times10^{-2}$** |
| Model | EfficientNetL | Gradient Boosting | **$6.93\times10^{-12}$** | **$5.04\times10^{-17}$** | **$4.92\times10^{-14}$** | **$1.06\times10^{-18}$** | **$2.67\times10^{-13}$** |
| Model | EfficientNetL | InceptionV3 | 1.00 | 1.00 | 1.00 | 1.00 | 1.00 |
| Model | EfficientNetL | MLP | **$7.33\times10^{-5}$** | **$7.53\times10^{-9}$** | **$3.26\times10^{-6}$** | **$7.00\times10^{-11}$** | **$2.20\times10^{-5}$** |
| Model | EfficientNetL | Random Forest | **$5.89\times10^{-11}$** | **$1.56\times10^{-17}$** | **$3.31\times10^{-13}$** | **$1.71\times10^{-17}$** | **$5.66\times10^{-13}$** |
| Model | EfficientNetL | SVM | **$4.48\times10^{-11}$** | **$4.19\times10^{-13}$** | **$1.72\times10^{-13}$** | **$1.59\times10^{-17}$** | **$8.91\times10^{-13}$** |
| Model | Gradient Boosting | InceptionV3 | **$1.10\times10^{-5}$** | **$9.46\times10^{-13}$** | **$1.27\times10^{-7}$** | **$4.67\times10^{-14}$** | **$3.49\times10^{-7}$** |
| Model | Gradient Boosting | MLP | **$1.04\times10^{-4}$** | **$7.04\times10^{-3}$** | **$1.63\times10^{-4}$** | **$4.79\times10^{-4}$** | **$1.11\times10^{-4}$** |
| Model | Gradient Boosting | Random Forest | 1.00 | 1.00 | 1.00 | 1.00 | 1.00 |
| Model | Gradient Boosting | SVM | 1.00 | 1.00 | 1.00 | 1.00 | 1.00 |
| Model | InceptionV3 | MLP | $2.66\times10^{-1}$ | **$9.89\times10^{-6}$** | **$2.33\times10^{-2}$** | **$1.68\times10^{-7}$** | $7.65\times10^{-2}$ |
| Model | InceptionV3 | Random Forest | **$5.30\times10^{-5}$** | **$7.04\times10^{-12}$** | **$5.61\times10^{-7}$** | **$3.66\times10^{-13}$** | **$8.22\times10^{-7}$** |
| Model | InceptionV3 | SVM | **$3.95\times10^{-5}$** | **$6.43\times10^{-11}$** | **$3.16\times10^{-7}$** | **$3.99\times10^{-13}$** | **$8.98\times10^{-7}$** |
| Model | MLP | Random Forest | **$8.83\times10^{-4}$** | $1.12\times10^{-1}$ | **$9.90\times10^{-4}$** | **$2.34\times10^{-3}$** | **$1.84\times10^{-4}$** |
| Model | MLP | SVM | **$9.42\times10^{-4}$** | **$4.94\times10^{-3}$** | **$5.77\times10^{-4}$** | **$3.85\times10^{-3}$** | **$3.61\times10^{-4}$** |
| Model | Random Forest | SVM | 1.00 | 1.00 | 1.00 | 1.00 | 1.00 |
| Sample Size | 24 | 48 | **$1.61\times10^{-3}$** | $1.60\times10^{-1}$ | **$2.70\times10^{-2}$** | $2.79\times10^{-1}$ | **$2.71\times10^{-2}$** |
| Sample Size | 24 | 100 | **$3.01\times10^{-13}$** | **$1.33\times10^{-5}$** | **$7.08\times10^{-10}$** | **$8.89\times10^{-6}$** | **$2.22\times10^{-11}$** |
| Sample Size | 24 | 248 | **$1.73\times10^{-13}$** | **$1.07\times10^{-9}$** | **$6.92\times10^{-12}$** | **$4.30\times10^{-8}$** | **$2.87\times10^{-12}$** |
| Sample Size | 24 | 500 | **$1.63\times10^{-13}$** | **$1.39\times10^{-11}$** | **$2.02\times10^{-12}$** | **$8.57\times10^{-9}$** | **$2.00\times10^{-13}$** |
| Sample Size | 24 | 1000 | **$9.04\times10^{-19}$** | **$1.26\times10^{-15}$** | **$3.50\times10^{-17}$** | **$4.16\times10^{-12}$** | **$8.81\times10^{-18}$** |
| Sample Size | 24 | 2000 | **$5.48\times10^{-21}$** | **$3.21\times10^{-18}$** | **$3.49\times10^{-19}$** | **$3.77\times10^{-12}$** | **$2.64\times10^{-23}$** |
| Sample Size | 24 | 4000 | **$2.70\times10^{-18}$** | **$7.80\times10^{-19}$** | **$3.85\times10^{-17}$** | **$8.70\times10^{-12}$** | **$2.53\times10^{-20}$** |
| Sample Size | 48 | 100 | **$1.02\times10^{-4}$** | **$3.52\times10^{-2}$** | **$1.32\times10^{-3}$** | $5.29\times10^{-2}$ | **$6.05\times10^{-5}$** |
| Sample Size | 48 | 248 | **$1.35\times10^{-8}$** | **$2.05\times10^{-6}$** | **$8.70\times10^{-8}$** | **$2.65\times10^{-5}$** | **$3.50\times10^{-8}$** |
| Sample Size | 48 | 500 | **$2.26\times10^{-9}$** | **$2.48\times10^{-8}$** | **$8.40\times10^{-9}$** | **$3.05\times10^{-6}$** | **$1.01\times10^{-9}$** |
| Sample Size | 48 | 1000 | **$1.28\times10^{-13}$** | **$4.51\times10^{-13}$** | **$1.04\times10^{-12}$** | **$6.82\times10^{-9}$** | **$2.86\times10^{-13}$** |
| Sample Size | 48 | 2000 | **$3.66\times10^{-16}$** | **$3.89\times10^{-16}$** | **$5.60\times10^{-15}$** | **$2.87\times10^{-9}$** | **$9.59\times10^{-19}$** |
| Sample Size | 48 | 4000 | **$5.81\times10^{-15}$** | **$3.78\times10^{-16}$** | **$3.98\times10^{-14}$** | **$1.42\times10^{-9}$** | **$5.30\times10^{-17}$** |
| Sample Size | 100 | 248 | **$7.95\times10^{-3}$** | $6.55\times10^{-2}$ | **$1.11\times10^{-2}$** | $1.05\times10^{-1}$ | **$2.76\times10^{-2}$** |
| Sample Size | 100 | 500 | **$3.38\times10^{-4}$** | **$1.72\times10^{-3}$** | **$4.90\times10^{-4}$** | **$9.65\times10^{-3}$** | **$5.20\times10^{-4}$** |
| Sample Size | 100 | 1000 | **$5.65\times10^{-7}$** | **$3.49\times10^{-6}$** | **$1.35\times10^{-6}$** | **$2.91\times10^{-4}$** | **$4.35\times10^{-6}$** |
| Sample Size | 100 | 2000 | **$1.25\times10^{-9}$** | **$7.06\times10^{-9}$** | **$5.33\times10^{-9}$** | **$5.03\times10^{-5}$** | **$5.35\times10^{-11}$** |
| Sample Size | 100 | 4000 | **$2.05\times10^{-10}$** | **$5.94\times10^{-10}$** | **$7.10\times10^{-10}$** | **$2.71\times10^{-6}$** | **$1.23\times10^{-11}$** |
| Sample Size | 248 | 500 | 1.00 | 1.00 | 1.00 | 1.00 | 1.00 |
| Sample Size | 248 | 1000 | 1.00 | 1.00 | 1.00 | 1.00 | 1.00 |
| Sample Size | 248 | 2000 | $6.43\times10^{-2}$ | $1.08\times10^{-1}$ | $1.17\times10^{-1}$ | 1.00 | **$1.40\times10^{-2}$** |
| Sample Size | 248 | 4000 | **$4.61\times10^{-4}$** | **$8.74\times10^{-3}$** | **$1.37\times10^{-3}$** | $8.24\times10^{-2}$ | **$1.05\times10^{-4}$** |
| Sample Size | 500 | 1000 | 1.00 | 1.00 | 1.00 | 1.00 | 1.00 |
| Sample Size | 500 | 2000 | 1.00 | 1.00 | 1.00 | 1.00 | $8.33\times10^{-1}$ |
| Sample Size | 500 | 4000 | **$3.76\times10^{-2}$** | $2.80\times10^{-1}$ | $6.20\times10^{-2}$ | $8.72\times10^{-1}$ | **$1.28\times10^{-2}$** |
| Sample Size | 1000 | 2000 | 1.00 | 1.00 | 1.00 | 1.00 | 1.00 |
| Sample Size | 1000 | 4000 | $1.76\times10^{-1}$ | $9.58\times10^{-1}$ | $2.66\times10^{-1}$ | 1.00 | **$3.81\times10^{-2}$** |
| Sample Size | 2000 | 4000 | 1.00 | 1.00 | 1.00 | 1.00 | 1.00 |

**Table 3.** Results of pairwise Mann–Whitney U tests with Bonferroni correction comparing classification performance between models and sample sizes across five evaluation metrics: F1 Score, AUC, Accuracy, Sensitivity, and Specificity. For model comparisons, tests were conducted across all combinations of model architectures, pooling across training sample sizes. For sample size comparisons, tests were conducted across all combinations of sample sizes, pooling across model architectures. Most comparisons yielded highly significant results ($p_{mwu} \ll 0.05$), suggesting that both model type and training sample size substantially influence classification outcomes.

| Reference | Sample Size | Modality | Model | AUC | Model Category | Classification Type |
|---|---|---|---|---|---|---|
| This Study | 24 | X-ray | EfficientNetL | 0.839 | Deep Learning-Based | Multi-class |
| Paul et al., Med Image Anal 2021 | 25 | X-ray | (DenseNet-121) + Autoencoder Ensemble | 0.79 | Deep Learning-Based | Multi-class |
| This Study | 48 | X-ray | EfficientNetL | 0.876 | Deep Learning-Based | Multi-class |
| This Study | 100 | X-ray | EfficientNetL | 0.925 | Deep Learning-Based | Multi-class |
| This Study | 248 | X-ray | EfficientNetL | 0.949 | Deep Learning-Based | Multi-class |
| This Study | 500 | X-ray | EfficientNetL | 0.972 | Deep Learning-Based | Multi-class |
| Zhang et al., Radiology 2020 | 500 | X-ray | CV19-Net | 0.94 | Deep Learning-Based | Multi-class |
| Hu et al., Med Phys 2022 | 812 | X-ray | VGG-19 | 0.987 | Deep Learning-Based | Multi-class |
| This Study | 1000 | X-ray | InceptionV3 | 0.987 | Deep Learning-Based | Multi-class |
| This Study | 2000 | X-ray | InceptionV3 | 0.991 | Deep Learning-Based | Multi-class |
| This Study | 4000 | X-ray | InceptionV3 | 0.996 | Deep Learning-Based | Multi-class |
| Chaddad et al., IEEE TNNLS 2022 | 5254 | X-ray | GMM-CNN + Random Forest | 0.9945 | Deep Radiomics-Based | Binary |
| Khan et al., Comput Intell Neurosci 2022 | 34579 | CT + X-ray | EffNet+VGG16+ELM | 0.991 | Deep Learning-Based | Multi-class |

**Table 4** Comparison of key characteristics including model type, sample size, model selection, and AUC performance for models developed in this study versus studies in the literature. Models developed in this study (EfficientNetL and InceptionV3) show a consistent increase in AUC with larger sample sizes, and outperform or match the performance of external benchmarks such as CV19-Net [7], VGG-19 [8], and hybrid models [10] at small, intermediate, and large sample sizes